# Order-Independent Texture Synthesis

Li-Yi Wei          Marc Levoy

Stanford University *

## Abstract

Search-based texture synthesis algorithms are sensitive to the order in which texture samples are generated; different synthesis orders yield different textures. Unfortunately, most polygon rasterizers and ray tracers do not guarantee the order with which surfaces are sampled. To circumvent this problem, textures are synthesized beforehand at some maximum resolution and rendered using texture mapping.

We describe a search-based texture synthesis algorithm in which samples can be generated in arbitrary order, yet the resulting texture remains identical. The key to our algorithm is a pyramidal representation in which each texture sample depends only on a fixed number of neighboring samples at each level of the pyramid. The bottom (coarsest) level of the pyramid consists of a noise image, which is small and predetermined. When a sample is requested by the renderer, all samples on which it depends are generated at once. Using this approach, samples can be generated in any order. To make the algorithm efficient, we propose storing texture samples and their dependents in a pyramidal cache. Although the first few samples are expensive to generate, there is substantial reuse, so subsequent samples cost less. Fortunately, most rendering algorithms exhibit good coherence, so cache reuse is high.

**Keywords:** Texture Synthesis, Texture Mapping, Graphics Hardware

## 1  Introduction

Texture synthesis techniques can be classified as either *explicit* or *implicit* [4, Chapter 2]; an explicit algorithm generates all the texture samples directly while an implicit algorithm answers a query about a particular sample without computing the whole texture. Most existing statistical texture synthesis algorithms [3, 5, 24, 14, 8, 16, 19, 9] are explicit. Since the value of each texture pixel is related to other pixels (such as histograms in [8, 16] and spatial neighborhoods in [5, 19, 9]), it is impossible to determine their values individually. On the other hand, most procedural texture synthesis techniques [4, 12, 13] are implicit since they allow texels to be evaluated independently.

Implicit texture synthesis offers several advantages over explicit texture synthesis. First, since only those texels that are actually used need to be evaluated, implicit methods are usually computationally cheaper than the explicit ones. There are several situations where only a portion of the texels in a texture are used: the textured object might be clipped by the viewing frustum, it might be partially occluded by other objects in the scene, or it might be faraway so that in a mipmapped implementation, only the low-resolution levels of the mipmap are required. Second, implicit methods consume less memory since they don't need to store the texture they are generating. Third, implicit methods are more flexible since they allow texture samples to be evaluated on demand and in any order. Unfortunately, implicit methods are usually less general than explicit ones. Due to the requirement of independent texel evaluation, implicit methods cannot use general statistical texture modeling based on inter-pixel dependencies.

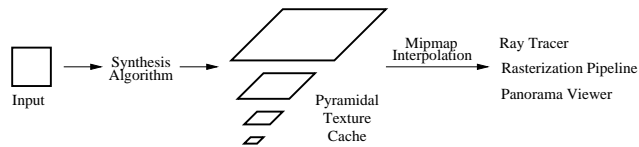

Figure 1: *Overview of our system. Given an input texture, our algorithm synthesizes a new texture and stores the synthesized samples in a pyramidal cache. Texture samples are computed on demand as requested by an external rendering system, such as a rasterization or ray tracing pipeline.*

The ideal texture synthesis algorithm would combine the advantages of both implicit and explicit techniques. The algorithm should be as general as statistical methods in that it can synthesize new textures simply from given examples. It should be as flexible as procedural methods in that it can allow textures to be evaluated on demand in any traversal order. In addition, different traversal orders should always yield identical results starting from the same initial conditions.

We describe a new algorithm that satisfies all the above requirements. This algorithm, termed *order-independent* texture synthesis, has constant time complexity for evaluating each output pixel where the constant depends only on the neighborhood and input image sizes. The algorithm allows texture samples to be generated in arbitrary order and at arbitrary resolution, yet the resulting texture remains identical. The algorithm is extended from previous multiresolution neighborhood-search texture synthesis algorithms [14, 19, 9] and generates textures with image quality comparable to those techniques.

An overview of our algorithm is shown in Figure 1. Given an input texture, our algorithm synthesizes new texture samples on demand as requested by an external rendering system, and stores the computed samples in a pyramidal cache. The rendering system can be a ray tracer, in which case our system functions as a texture shader; a rasterization pipeline, in which case our system functions as a mipmap texture cache; or a panorama viewer, in which case our system functions as a texture decompresser. Since our algorithm generates a texture on the fly rather than storing or transmitting the whole texture, it may require less storage or network bandwidth for managing large textures.

The rest of the paper is organized as follows. In Section 2, we review previous work. In Section 3, we describe our algorithm. In Section 4, we examine the algorithm behavior and demonstrate synthesis results. In Section 5, we simulate our algorithm with several texture mapping benchmarks. In Section 6, we conclude the paper and discuss limitations and future work.

## 2 Previous Work

**Statistical Synthesis.** Statistical synthesis algorithms model textures as a set of statistical features, and generate new textures by matching the statistics between the new texture and a given input sample. Possible texture statistics include histograms [8, 16], cross-scale pyramid dependencies [3], Markov Random Fields [14, 24], and searching spatial neighborhoods [5, 19, 9]. These algorithms are general and can generate many different textures based on input samples. They have also been extended to a variety of interesting applications such as texture transfer [1, 6], image analogies [9], and synthesis over surfaces [18, 20]. However, because these methods impose statistical dependencies between texture samples, it is impossible to evaluate only a subset of the samples while guaranteeing that these evaluated samples remain identical with respect to different synthesis orders. For example, histograms [8, 16] need to be computed from all texture samples, and neighborhood-searching [5, 19, 9] will produce different results if adjacent samples are synthesized in different orders. Although this problem can be partially addressed by generating textures in patches rather than individual pixels [23, 15, 6], for certain textures these approaches can produce visible boundary discontinuities at adjacent patches.

**Procedural Synthesis.** Procedural synthesis algorithms simulate the texture formation process using specialized procedures. Example procedures include noise [13], reaction-diffusion [17, 21], and cellular texturing [22, 11]. By using specialized procedures these techniques can be highly efficient and some of them allow texels to be evaluated independently from each other. However, since different textures may require different procedures, these algorithms are less general than statistical methods. In addition it can be difficult to tune the parameters for those procedures to achieve the desired visual appearance of the result texture.

## 3 Algorithm

Our algorithm is inspired by previous multiresolution neighborhood-search texture synthesis methods [19, 9, 18, 20]. Given a sample texture image, these algorithms synthesize a new texture pixel by pixel from lower to higher resolutions in a certain order, which can be scanline [19, 9], spiral [19, constrained synthesis], random [20], or surface sweeping [18]. To determine the value of a particular output pixel, its spatial neighborhood is compared against all possible neighborhoods from the input image. The input pixel with the most similar neighborhood is then assigned to the output pixel. This process is repeated for every output pixel at every pyramid level until the whole texture is assigned. Despite the simplicity of neighborhood searching, these algorithms have been demonstrated to work well over a wide variety of textures.

Unfortunately, these methods are not order-independent; different synthesis orders produce different synthesis results. This is because their search neighborhoods always include the most recently synthesized results, causing cyclic dependencies among texture pixels, as shown in Figure 2. To allow order-independent synthesis, we need to remove these cyclic neighborhood dependencies. We achieve this by a very simple idea: instead of overwriting the old pixel values with new results, we keep the old and new values

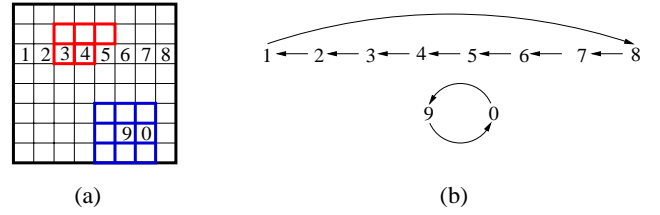

(a)                    (b)

Figure 2: *Cyclic neighborhood dependency. (a) An 8×8 texture with several numbered pixels. Assume that pixels 1 to 8 are generated by the red causal neighborhood, and pixels 0 and 9 are generated by the blue symmetric neighborhood. (b) The neighborhood dependency among those pixels is cyclic, preventing order-independent synthesis. Note that pixel 1 depends on pixel 8 since the neighborhood is toroidal around the boundary [19].*

*in separate images* and use only the old values in the search process. This idea is inspired by image convolution. When convolving image $X$ by a filter kernel to produce image $Y$, each pixel in $Y$ is computed by convolving the kernel with the "old" pixels in $X$ rather than the new pixels in $Y$. As a result, there are no cyclic dependencies among pixels in $Y$, and we can compute their values in any order. However, unlike convolution where only one old value is retained, texture synthesis may generate and retain several old values for each pixel, by performing multiple synthesis passes. For example, both constrained synthesis [19] and random order synthesis [20] uses a first pass to generate an initial guess, and improves the result in subsequent passes. In our algorithm, we keep multiple generations for each pixel, where each generation corresponds to a separate pass in [19, 20].

---

**function** $C \leftarrow$ SynthesizePixel($G_a, L, p, m$)
1   **if** CacheHit($L, p, m$)
2       return CacheValue($L, p, m$);
3   **else**
4       $N_s \leftarrow$ BuildOutputNeighborhood($L, p, m$);
5       $N_a^{best} \leftarrow$ null;    $C \leftarrow$ null;
6       **loop** through all pixels $p_i$ of $G_a(L)$
7           $N_a \leftarrow$ BuildInputNeighborhood($G_a, L, p_i$);
8           **if** Match($N_a, N_s$) > Match($N_a^{best}, N_s$)
9               $N_a^{best} \leftarrow N_a$;    $C \leftarrow G_a(L, p_i)$;
10      AddCacheEntry($L, p, m, C$);
11      **return** CacheValue($L, p, m$);

**function** $N_s \leftarrow$ BuildOutputNeighborhood($L, p, m$)
12  $N_s \leftarrow$ null;
13  **foreach** $(L_n, p_n, m_n) \in$ Neighborhood($L, p, m$)
14      % *note:* $(L_n, m_n) \prec (L, m)$
15      **if** CacheHit($L_n, p_n, m_n$)
16          $N_s \leftarrow N_s \oplus$ CacheValue($L_n, p_n, m_n$);
17      **else**
18          $C \leftarrow$ SynthesizePixel($G_a, L_n, p_n, m_n$);
19          $N_s \leftarrow N_s \oplus C$;
20  **return** $N_s$;

Table 1: *Pseudocode of order-independent texture synthesis. See Table 2 for the meanings of the symbols.*

Our algorithm contains three major extensions over [19, 9, 18, 20]. First, we retain several generations for each output pixel, as we have shown earlier. Second, we restrict the output neighborhood to contain only pixels from lower pyramid resolutions and earlier generations. This removes the cyclic dependencies among output pixels. Specifically, for pixel located at (level $L_1$, genera-tion $m_1$), its neighborhood can contain pixels only from $(L_2, m_2)$ where $[L_2 < L_1]$ or $[L_2 = L_1$ and $m_2 < m_1]$. The first condition $[L_2 < L_1]$ implies pixels at lower resolutions and the second condition $[L_2 = L_1$ and $m_2 < m_1]$ implies pixels at the same resolution but earlier generations. We use the lexically-smaller operator $\prec$ to represent such relationship, and allow $(L_2, m_2)$ belongs to the neighborhood of $(L_1, m_1)$ only if $(L_2, m_2) \prec (L_1, m_1)$.

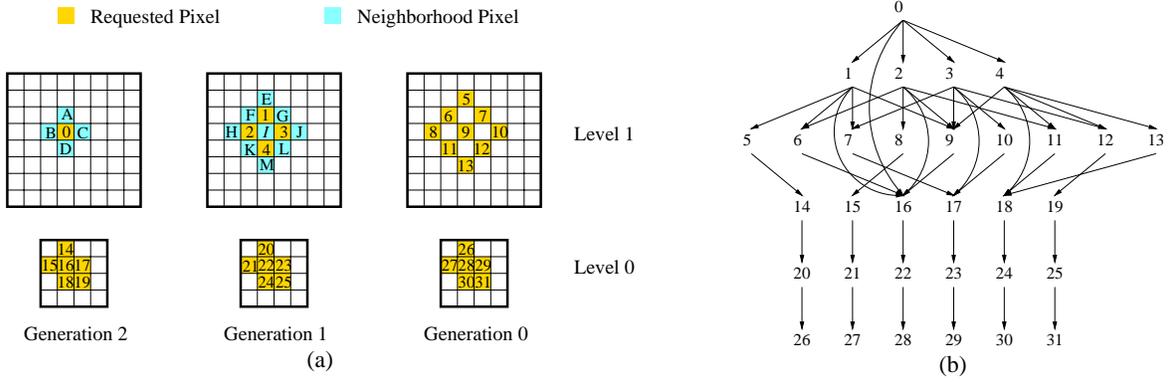

Figure 3: *Synthesizing one pixel using our algorithm. The pyramid cache consists of two resolutions and three generations, as shown in (a). Assume a pixel, marked 0, is requested, and we synthesize it using a small neighborhood, containing 4 adjacent pixels at the same resolution and 1 pixel from the lower resolution. In this example, we use this small 5-pixel neighborhood for clarity; for real synthesis, we usually use much larger neighborhoods (usually 5×5). In previous algorithms, the neighborhood of pixel 0 contained pixel A, B, C, D, and 16. This causes cyclic dependencies between 0 and A to D. In our algorithm, the neighborhood contains pixels 1 to 4 instead of A to D, and therefore removes the cyclic dependencies. Similarly, pixels 1 to 4 depend on pixels 5 to 13 rather than pixels E to M. For neighborhood pixels at lower resolutions, we always use the latest generation; for example, pixel 1 depends on 16 rather than 22. For pixels 5 to 13, the neighborhood can contain pixels only from lower resolutions since there is no pixel at further lower generations. Similarly, for pixels 14 to 19, the neighborhood can contain pixels only from lower generations since there is no pixel at further lower resolutions. Finally, pixels 26 to 31 cannot depend on any other pixels since they locate at the lowest resolution and generation. They are initialized at the beginning of the algorithm to a noise image and do not require computation. The complete dependency graph among all pixels is shown in (b). The algorithm computes the pixels by traversing the dependency graph in a depth first order.*

| Symbol | Meaning |
|---|---|
| $G_a$ | Input texture pyramid |
| $p_i$ | An input pixel |
| $p$ | An output pixel |
| $L$ | pixel level |
| $m$ | pixel generation/iteration |
| $C$ | pixel color |
| $N(p)$ | Neighborhood around the pixel $p$ |
| $G(L)$ | $L$th level of pyramid $G$ |
| $CacheValue(L, p, m)$ | cached color $C$ for pixel $(L, p, m)$ |
| $\prec$ | lexically smaller |

Table 2: *Table of symbols for the pseudo-code in Table 1. $(L_2, m_2) \prec (L_1, m_1)$ if $[L_2 < L_1]$ or $[L_2 = L_1$ and $m_2 < m_1]$.*

Third, instead of computing all pixels sequentially from lower to higher resolutions and storing them in the output pyramid, we evaluate them on the fly and store the computed values in a pyramidal cache. This allows individual texture samples to be evaluated on demand without computing the whole texture. The cache also allows previously computed pixels to be reused later. In the beginning of the algorithm, the cache is empty and every requested pixel needs to be computed. However, as the cache gradually fills up, previously computed pixels may be requested again, and they can be found in the cache without any computation. The actual extent of cache reuse depends on the rendering algorithm that drives these requests. Fortunately, since most practical rendering algorithms exhibit good coherence, cache reuse is typically high.

Our complete algorithm combining these key ideas is shown in

Table 1. The cache consists of entries $(L, p, m, C)$, where $L$ is the pixel level, $p$ is the pixel location $(x, y)$, $m$ is the generation number, and $C$ is the pixel color. The portion $(L, p, m)$ is the cache tag and $C$ is the cache value. To synthesize a specific pixel $(L, p, m)$ (**SynthesizePixel**), we first check if the pixel is in the cache. If so, no computation is required and the cache entry is returned. Otherwise, we build the neighborhood around $(L, p, m)$ and search for the best match from the input pyramid $G_a$. The code for neighborhood searching (lines 4 through 9 in Table 1) is very similar to [19], except that we use different ways to build input and output neighborhoods. The input neighborhood is built as in [19] (**Build-InputNeighborhood**), but the output neighborhood is built from the cache rather than an output pyramid (**BuildOutputNeighborhood**).

The function **BuildOutputNeighborhood** works as follows. For each pixel $(L_n, p_n, m_n)$ in the neighborhood of the output pixel $(L, p, m)$, we first check if it is in the cache. If so, we add it directly to the output neighborhood $N_s$. Otherwise, we call **SynthesizePixel** recursively to compute its value and add the computed value to $N_s$. Note that we require each $(L_n, m_n)$ to be lexically smaller than $(L, m)$, so that the Neighborhood$(L, p, m)$ can contain only pixels from lower resolutions, as well as pixels from the same resolution which are generated in earlier iterations. Because of this, the dependencies of the pixels form an acyclic graph and the mutual recursive calls between **SynthesizePixel** and **BuildOutputNeighborhood** are guaranteed to terminate, unless the cache is too small to simultaneously hold all pixels required to compute $(L, p, m)$. By analyzing the algorithm, it can be easily shown that each output pixel depends on a fixed number of neighborhood pixels at lower resolutions and earlier generations. An example for synthesizing one pixel using this algorithm is shown in Figure 3.

**Initialization.** We initialize the lowest resolution/generation $(L_{min}, m_{min})$ of the cache by copying pixels randomly from the lowest resolution of the input pyramid. In other words, each output pixel at $(L_{min}, m_{min})$ is assigned a random value from the input. This initialization completely determines the synthesis re-sult since each output pixel is determined only by these pixels at $(L_{min}, m_{min})$. (This can be shown by recursively expanding the dependency graph for each pixel, following the mutual calls between **SynthesizePixel** and **BuildOutputNeighborhood**.) This initialization can be implemented by either permanently storing pixels $(L_{min}, m_{min})$ in the cache, or by using a pseudo-random number table to choose the random values on the fly as implemented in Perlin noise [13]. In our experience there is no general restriction about this initialization process. The only requirement is to avoid initializing the lowest resolution with a constant color, which generates a constant colored texture due to the order-independency of the algorithm.[1]

In addition to pixels at $(L_{min}, m_{min})$, we can also optionally add pixels at higher resolutions/generations during initialization. This can be useful in situations where we want to "fix" the values of certain pixels. For example, in constrained synthesis [5, 19] we may want to replace a foreground object from a texture background, while keeping the background unchanged. This can be done by adding a small set of background pixels around the object into the cache during initialization, as follows. We acquire the boundary pixels located at several levels from the original pyramid, and add them to the highest generation at the corresponding cache pyramid levels.

## 4 Synthesis Results

In this section, we demonstrate several aspects of our algorithm's performance. We begin by comparing the quality of synthesized textures with previous methods. We then characterize the cache performance under access patterns with different amount of coherence.

**Synthesis Quality.** In Figure 5, we compare the results generated by our algorithm and earlier methods [19, 20]. We use similar parameters for both versions of algorithms: Gaussian pyramid with 4 levels, a neighborhood of size $5\times5$ with 2 levels, three passes for [19, 20] and three generations for our method. As shown, the algorithm generates results with quality comparable to earlier methods. We believe this is because our use of multiple generations achieves similar functionality as the multiple passes in previous methods.

**Spatial Cache Coherence.** If every request for a texture pixel resulted in a complete evaluation of the dependency tree as depicted in Figure 3, then our algorithm would be very slow. Decent performance depends critically on these requests reusing intermediate results that we store in the pyramidal cache. The amount of reuse depends on the coherence of the request patterns. There are two kinds of access coherence: spatial coherence and temporal coherence. An access pattern has more spatial coherence if its requests are of nearby texture locations. An access pattern has more temporal coherence if recently requested pixels are likely to be reused again in the near future.

To begin our analysis, let us consider several access patterns with different amounts of spatial coherence as shown in Figure 7. For each request pattern, we show the content of the cache for pixels at different levels and generations. Figure 7 (a) shows the cache footprint for synthesizing one pixel. We can see that many cache pixels may be touched in order to synthesize one pixel. However, the cost of generating multiple cache pixels can be amortized by requesting multiple pixels, as shown in Figure 7 (b) and (c). The amount of amortization depends on the spatial coherence of specific access patterns. For example, both (b) and (c) synthesize the same number of output pixels. However, since (c) has a circular request pattern, its cache footprint is more coherent and much fewer pixels are touched in the cache compared to (b).

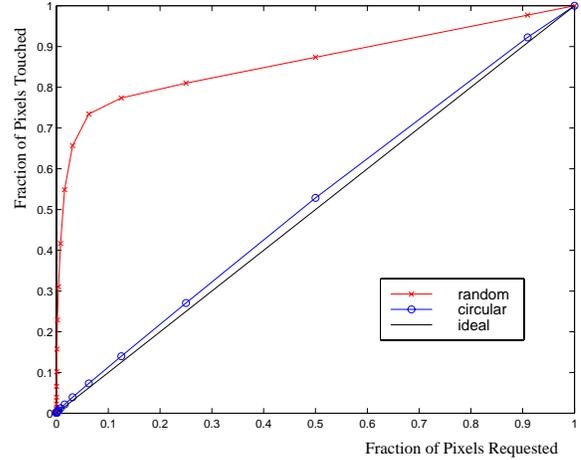

**Figure 4:** *Texture cache usage. This diagram illustrates what fraction of the cache will be touched for different input requests. The horizontal axis indicates the fraction of pixels requested for a $512\times512$ texture, and the vertical axis indicates the fraction of pixels touched in the corresponding texture cache, large enough to hold all output pixels. Two different request patterns are shown. The blue curve indicates a circular access pattern and the red curve indicates a random access pattern (Figure 7). The black line indicates the ideal linear behavior. The vertical intercept of the red and blue curves at the left is about 0.0011 (1148 pixels as in Figure 7), indicating the large (although constant) footprint associated with synthesizing one pixel.*

To quantitatively measure the effect of spatial coherence on our algorithm's performance, we vary the amount of requested pixels for these two patterns, and plot the corresponding number of synthesized cache pixels in Figure 4. The ideal case is a linear relationship between the number of requested and touched pixels. As shown, the circular pattern offers near optimum behavior and is almost linear, whereas the random pattern has worse performance and requests more cache pixels. These two patterns demonstrate the best and worst case cache behavior for the particular search neighborhood size and shape used in Figure 7. For these neighborhoods, renderings of real scenes will have spatial coherence bounded between the two extremes shown in Figure 4.

**Temporal Cache Coherence.** As in most cache-based algorithms, the size of the cache plays a critical role in its performance. When the cache in our algorithm is large, it can hold all the computed pixels and its performance is determined by the spatial coherence of the request patterns, as we have seen in Figure 4. However when the cache is small, it cannot hold all pixels, and some pixels may be computed multiple times. As a result, the performance of small caches is affected by not only the spatial coherence but also the temporal coherence of the request patterns.

To analyze the effect of temporal coherence on performance for caches of different sizes, we generate artificial access patterns with different amount of temporal locality, and feed them into our algorithm with different cache sizes. We generate these access patterns in tiled rasterization order [7], as follows. We partition a large $512\times512$ texture into tiles of different sizes, and synthesizes the texture tiles one by one in a scanline order. Within each tile, we traverse the tile pixels in a scanline order. The tile size controls the temporal locality of the access pattern. When the tile size is too large or too small, the temporal coherence is low since the rasterization order contains long scanlines. When the tile has medium size, the temporal locality is high since pixels within the same tile may have significant overlaps in their neighborhoods.

---

[1] If the lowest resolution contains a single pixel, we conduct the random initialization at a higher resolution (usually at size $4\times4$) and generate the further lower resolutions by filtering/downsampling this randomized resolution.

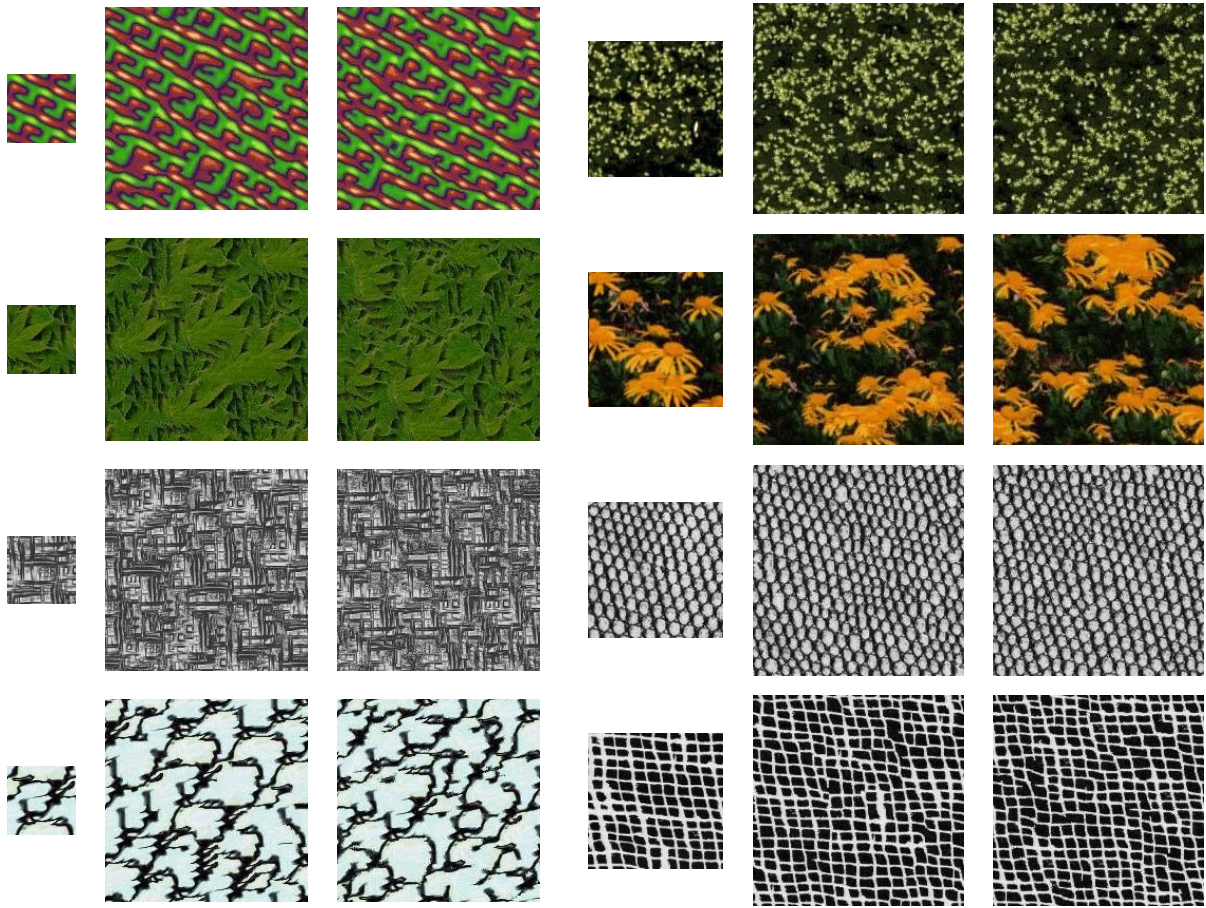

Figure 5: *Quality comparison between order-independent synthesis and previous methods. For each group of images, the left is the original texture, the middle is generated by [19], and the right is generated by order-independent synthesis.*

We measure the performance of our algorithm under different combinations of cache and tile sizes, and the performance of our algorithm is plotted in Figure 8. As shown, when the cache is bigger than 32K, the algorithm is insensitive to tile size since the majority of pixels can reside in the cache. However when the cache is smaller, the algorithm performs better on medium tile sizes (between $8{\times}8$ and $128{\times}128$) rather than small and large tile sizes. This indicates that at small cache sizes, the temporal coherence of the requests can significantly determine the performance of our algorithm.

## 5   Texture Mapping Performance

We now demonstrate how our algorithm works in practice. In particular, we have profiled our algorithm using several texture mapping benchmarks with different texture characteristics and triangle sizes. We collect GL traces from these benchmarks, run the traces through a hardware simulator, and from the simulated fragment generator we intercept texture access calls. The recorded texture accesses are then fed into our algorithm for simulation.

Our algorithm garners advantages over synthesizing the entire texture ahead of time in a variety situations which cause only part of the texture to be accessed. For example, the textured object might be clipped by the viewing frustum, it might be partially occluded by other objects in the scene, or it might be faraway, so that only a low resolution texture is required.

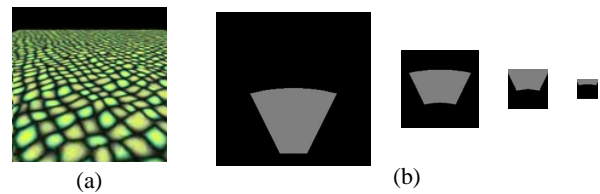

(a)                                      (b)

Figure 6: *A textured polygon in perspective. (a) The perspective view of a $512{\times}512$ square polygon tessellated into $64{\times}64$ tiles. (b) The requested texels (shown in gray color) at 4 mipmap levels. The unexpected curves in the mipmap levels in (b) arise from the approximate method used to determine which mipmap level to request for each rendered pixel in (a). The screen resolution is $512{\times}512$ and the mipmap filtering is trilinear interpolation.*

In the rest of this section, we first describe the characteristics of each benchmark scene, and we summarize the simulation results for all scenes in Figure 9.

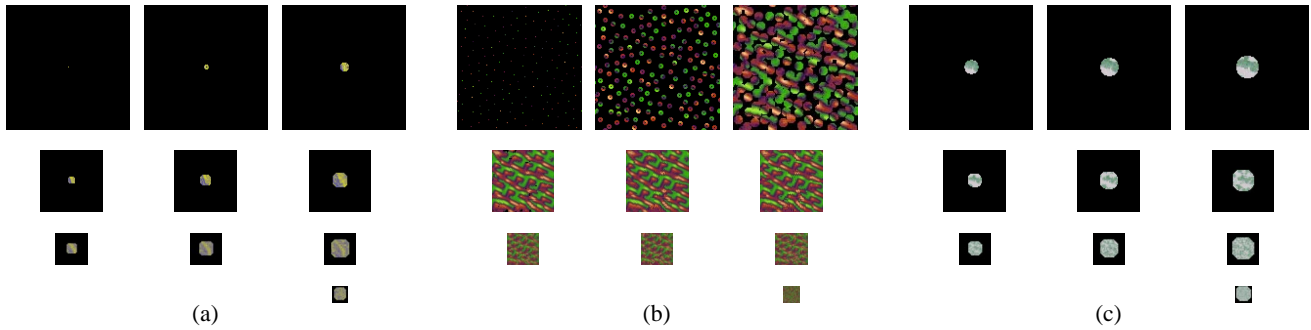

(a)                     (b)                     (c)

Figure 7: *Cache access footprints for various request patterns: (a) a single pixel, (b) a set of uniform random pixels generated by a poisson disk process, and (c) a circular request pattern. Within each group, the images show the contents of the cache at different generations and different pyramid levels, with higher resolution on the top and later generation on the left. We use 4 pyramid levels and 3 generations except the lowest resolution where only one generation (the initial value) is used, and a circular neighborhood of size 5×5 with 2 levels. Image sizes are 128x128, 64x64, 32x32, and 16x16, respectively, from top to bottom. The total number of footprint pixels in (a) is 1148.*

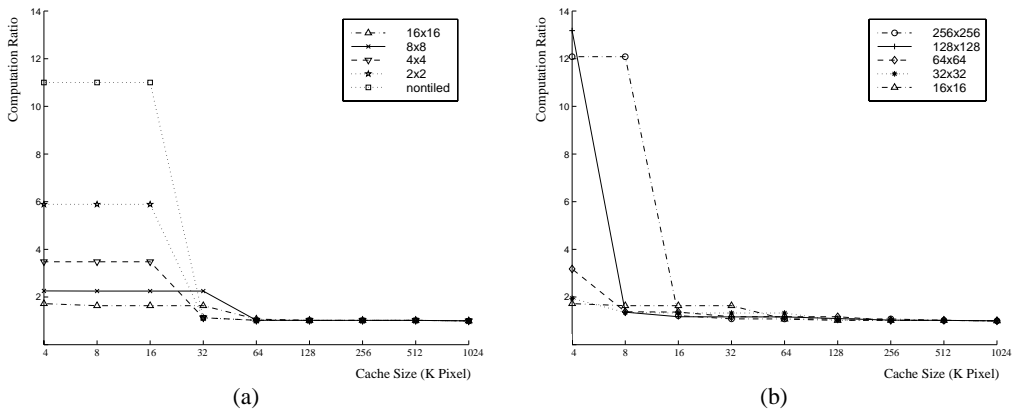

(a)                                   (b)

Figure 8: *Effect of cache size on our algorithm. We measure the cache performance by* computation ratio, *which we define to be the average number of computed texels per requested texel. The performance is better when the computation ratio is lower since less computation is required. To exploit temporal cache coherence, we generate requests in tiled rasterization order with different tile sizes. We feed these requests into our algorithm, and plot the resulting computation ratio for different combinations of tile and cache sizes. (a) Small to medium tile sizes. (b) Medium to large tile sizes. Tile sizes are given in pixels.*

- **Single Polygon.** Figure 6 shows a large textured polygon with size 512×512, viewed in perspective. The polygon is tessellated into 64×64 tiles and covered by a large 512×512 texture pyramid with 4 levels. The viewpoint is chosen to be close enough so that all 4 levels of the mipmap are accessed. As shown in Figure 6 (b) the texture access pattern is coherent across different mipmap levels. In this example, we explore what happens when only a portion of the texture is visible. Since the polygon is clipped by the viewing frustum, only 19 percent of the texels are requested by the rasterizer. Our algorithm synthesizes 23 percent of all cache pixels. Although this is slightly larger than 19 percent, it is still 4 times faster than computing the whole texture.

- **Quake.** Figure 10 shows a frame from the OpenGL port of the video game Quake. This application is essentially an architectural walkthrough with software visibility culling. The scene contains mainly large polygons with repeating textures. We choose this benchmark to see how rasterizing large polygons may degrade cache coherence, as shown in Figure 8. In the first part of this experiment (Figure 10 (a)), we synthesize the 64×64 lava texture on the floor from a small 48×48 crop (Figure 10 (c, e)). The 64×64 texture size is specified in the original quake benchmark. Because the texture is repeated and viewed in perspective under different distances, 6 levels of the texture mipmap are accessed.

The Quake benchmark uses many small textures to reduce texture load; although this works well for regular patterns such as brick walls, small textures may introduce unnatural visual repetitions for stochastic textures such as the lava. Our algorithm can address this problem by synthesizing large textures on the fly from a small input. We demonstrate this in the second part of this experiment (Figure 10 (b)). We synthesize the lava texture with size 512×512 from the same 48×48 crop (Figure 10 (d)), and eliminate the majority of the unnatural repetitions in the rendering. Although the texture is large, only 44 percent of texels are requested by the rasterizer and our algorithm computes only 48 percent of all cache pixels. This fraction will vary, of course, depending on the path the user takes through the model.

- **QuickTime VR.** Figure 11 is a frame of an OpenGL-based QuickTime VR viewer looking at a panorama of Mars. This large panorama (8K×1K) is divided into 256×256 texture tiles and is mapped onto a cylinder made of tall, skinny triangles. Because the panorama is viewed in high screen reso-lution, only the highest resolution of the mipmap is accessed. For the purpose of this experiment, we choose one 256×256 texture tile that contains repetitive patterns and replace the central 224×224 region of it by synthesizing a new texture on demand from a small 64×64 crop taken from the same region (Figure 11 (c, d, e)). This essentially compresses the original 224×224 texture into a small 64×64 crop. In addition, since the bottom of the panorama is partially outside the viewing frustum, part of the texture is never referenced and therefore not synthesized, as shown by the black region in Figure 11 (d).

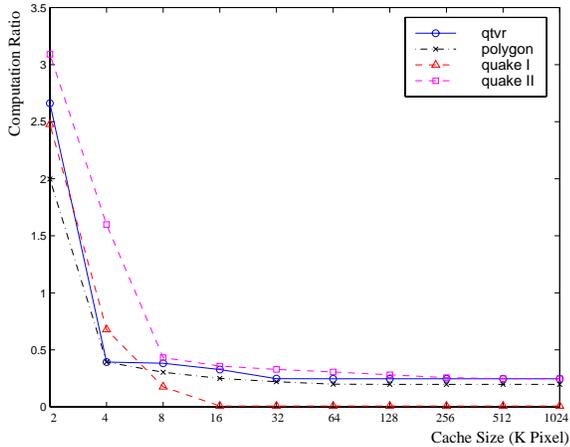

Figure 9: *Texture synthesis statistics for several benchmark scenes: qtvr (Figure 11), polygon (Figure 6), quake I with a 64×64 texture, and quake II with a 512×512 texture (Figure 10). The horizontal axis indicates cache sizes and the vertical axis indicates the computation ratio (average number of computed pixels per requested pixel). Note that the vertical axis can go as high as 1148 if only a single pixel is requested (Figure 7).*

**Results.** In Figure 9, we plot the computation ratio versus different cache sizes for all three benchmarks. The algorithm performs reasonably well at small cache sizes (between 2K to 8K), and the computation ratio drops as the cache size increases. The performance remains roughly constant after the cache size reaches 16K, indicating that a smaller cache size is sufficient to hold the working set for these scenes. Note that if only a single pixel is requested, the computation ratio will be very high (around 1148 pixels). Fortunately, since practical benchmarks usually contain sufficient access coherence, our algorithm performs fairly well: the computation ratio is usually below 1 and never exceeds 3.5 for these benchmarks.

We now describe detailed performance of different benchmarks. The QuickTime VR and single polygon benchmarks have similar performance since they both render large textures with medium-sized triangles. On the other hand, the two Quake benchmarks perform differently. At small cache sizes (2k and 4K), Quake II has higher computation ratio than the other three benchmarks since the scene contains a large texture over large triangles, causing the texture to be traversed in long scanlines, reducing the temporal coherence (as demonstrated in Figure 8). However, large triangles are not a problem for Quake I since it uses a small texture. As the cache size increases the small texture of Quake I can fit entirely in the cache. Because the texture is repeated over the floor, it is requested multiple times and the computation ratios drop to almost zero (but not zero since the texture needs to be computed at least once). Of course, since the texture in Quake I is small, one can argue that it could as easily be stored as synthesized, so the advantage of synthesizing it is modest. On the contrary, the other three bench-marks have large textures which are accessed only once. Since new pixels are always computed for each request, the computation ratio remains constant around 0.25 even at large cache sizes. This constant is determined by the neighborhood sizes and amount of neighborhood overlapping between adjacent requests.

# 6 Conclusions and Future Work

In this paper, we have presented and analyzed a new algorithm for order-independent texture synthesis. The algorithm allows texture samples to be generated in arbitrary order on demand, yet the resulting texture always looks the same, and the algorithm has comparable computational expense to previous neighborhood-based texture synthesis algorithms. It is also storage efficient due to its use of a pyramidal cache. We also demonstrate that small caches are sufficient by analyzing our algorithm through different texture mapping benchmarks.

There are several possible directions for future work. Since our algorithm has the flexibility of a procedural texture shader, it can be implemented in a shading language and integrated with a ray tracing package. This could make statistical texture synthesis more useful to animators, who are more accustomed to procedural shaders. Our algorithm can also be used as a texture decompresser for a software panorama viewer such as QuickTime VR. This would substantially reduce the storage space and transmission time for viewing panoramas containing large textured regions. Finally, it can be combined with a hardware rendering pipeline, functioning in that context like a mipmap texture cache. We believe that many observations made in previous texture caching papers [7, 2, 10] could be applied in our algorithm, which has similar (but not identical) characteristics. For example, both texture caching and our algorithm perform better when the access pattern exhibits good coherence.

Another possible future direction would be to combine our algorithm with patch-based texture synthesis algorithms [23, 15, 6]. It has been shown that both pixel-based synthesis [19, 9, 18, 20] and patch-based synthesis have limitations and work well on different textures. For example, pixel-based approaches tend to fail on textures with meaningful textons while patch-based approaches tend to produce visible boundary discontinuities at adjacent texture patches. Our approach, being extended from pixel-based algorithms, has limitations similar to [19, 20]. However, we can combine the advantages from both synthesis methods by using patch-based algorithm to pre-compute a texture partition, and during run time using our algorithm to fill in the gaps/overlaps between adjacent texture patches. Since both patch-based synthesis and our algorithm allow independent texture query, such a hybrid would support real time texture synthesis with image quality better than both patch and pixel based methods. A potential future direction would be to design a hardware or software system that efficiently implements this combination.

## Acknowledgments

We would like to thank Kekoa Proudfoot and Ian Buck for providing the glsim code, Ravi Ramamoorthi and James Davis for commenting on early drafts of the paper, and members of the Stanford Graphics Group for their support. This research is sponsored by Intel, Interval, and Sony under the Stanford Immersive Television Project.

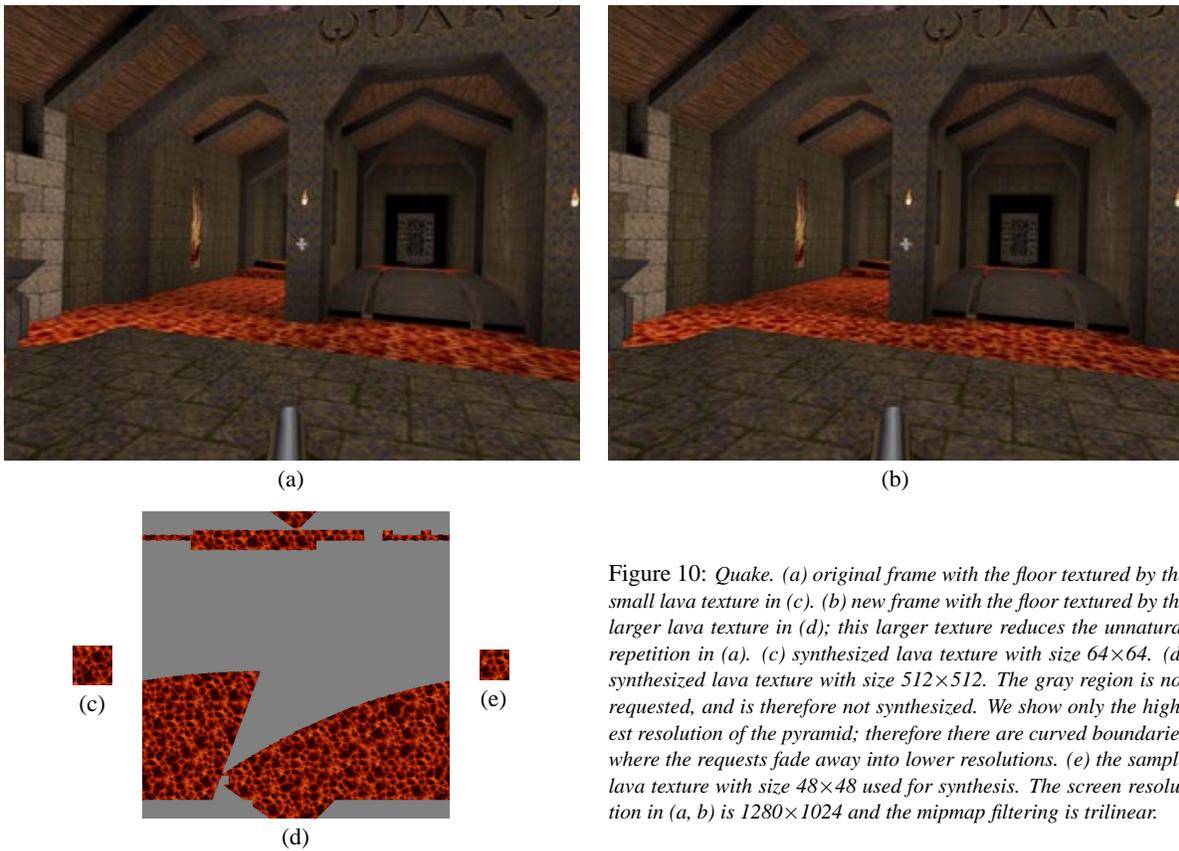

(a)

(b)

(c)

(e)

(d)

Figure 10: *Quake. (a) original frame with the floor textured by the small lava texture in (c). (b) new frame with the floor textured by the larger lava texture in (d); this larger texture reduces the unnatural repetition in (a). (c) synthesized lava texture with size 64×64. (d) synthesized lava texture with size 512×512. The gray region is not requested, and is therefore not synthesized. We show only the highest resolution of the pyramid; therefore there are curved boundaries where the requests fade away into lower resolutions. (e) the sample lava texture with size 48×48 used for synthesis. The screen resolution in (a, b) is 1280×1024 and the mipmap filtering is trilinear.*

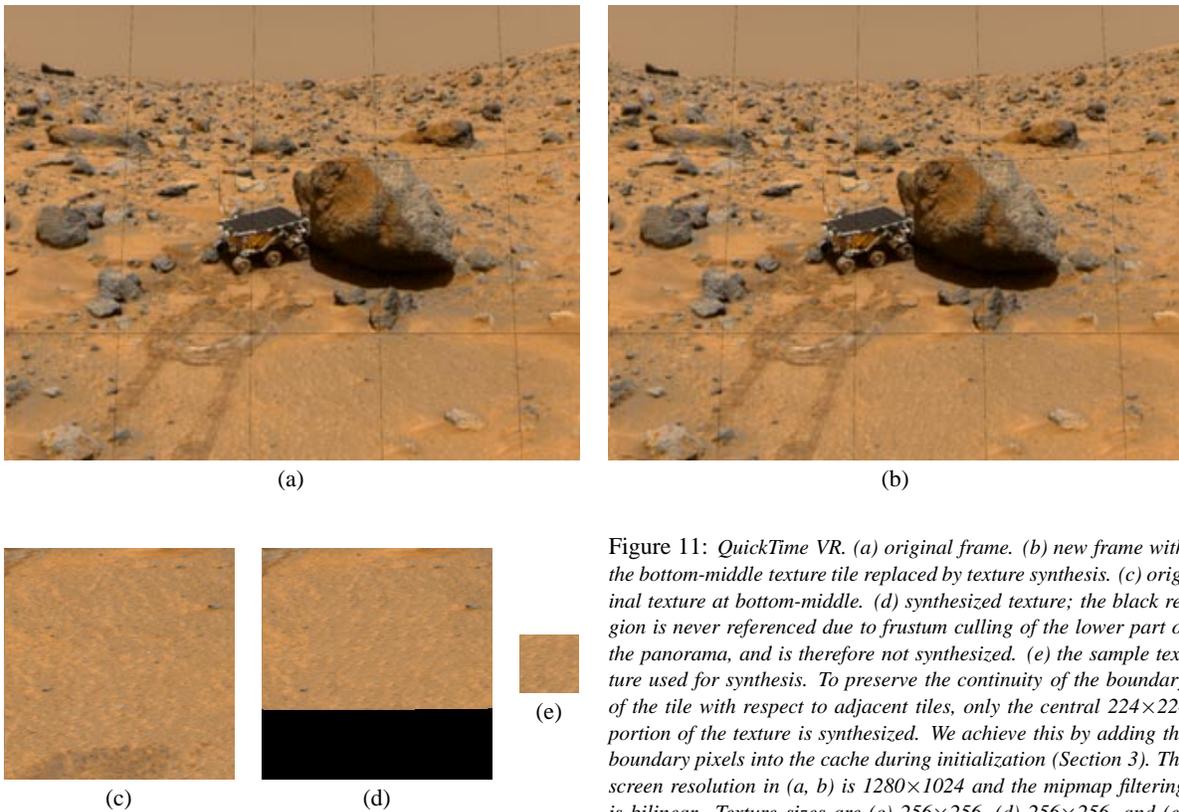

(a)

(b)

(c)

(d)

(e)

Figure 11: *QuickTime VR. (a) original frame. (b) new frame with the bottom-middle texture tile replaced by texture synthesis. (c) original texture at bottom-middle. (d) synthesized texture; the black region is never referenced due to frustum culling of the lower part of the panorama, and is therefore not synthesized. (e) the sample texture used for synthesis. To preserve the continuity of the boundary of the tile with respect to adjacent tiles, only the central 224×224 portion of the texture is synthesized. We achieve this by adding the boundary pixels into the cache during initialization (Section 3). The screen resolution in (a, b) is 1280×1024 and the mipmap filtering is bilinear. Texture sizes are (c) 256×256, (d) 256×256, and (e) 64×64.*

# Order-Independent Texture Synthesis

Category: Research

## Abstract

Search-based texture synthesis algorithms are sensitive to the order in which texture samples are generated; different synthesis orders yield different textures. Unfortunately, most polygon rasterizers and ray tracers do not guarantee the order with which surfaces are sampled. To circumvent this problem, textures are synthesized beforehand at some maximum resolution and rendered using texture mapping.

We describe a search-based texture synthesis algorithm in which samples can be generated in arbitrary order, yet the resulting texture remains identical. The key to our algorithm is a pyramidal representation in which each texture sample depends only on a fixed number of neighboring samples at each level of the pyramid. The bottom (coarsest) level of the pyramid consists of a noise image, which is small and predetermined. When a sample is requested by the renderer, all samples on which it depends are generated at once. Using this approach, samples can be generated in any order. To make the algorithm efficient, we propose storing texture samples and their dependents in a pyramidal cache. Although the first few samples are expensive to generate, there is substantial reuse, so subsequent samples cost less. Fortunately, most rendering algorithms exhibit good coherence, so cache reuse is high.

**Keywords:** Texture Synthesis, Texture Mapping, Graphics Hardware

## 1 Introduction

Textures are employed to represent surface details in synthetic scenes without adding geometric complexity. A common way to acquire textures is by texture synthesis. Existing texture synthesis techniques can be classified as either *implicit* or *explicit* [Ebert et al. 1998, Chapter 2]; an implicit method computes only required texels on demand while an explicit method always computes the entire texture. For example, Perlin noise is an implicit method while histogram matching [Heeger and Bergen 1995] is an explicit method. Implicit methods are usually more flexible and storage efficient than explicit ones since they compute only needed texels on the fly without storing the entire texture. On the other hand, explicit methods are often more general than implicit ones since they can model many different textures based on image statistics. In addition, explicit methods are easier to use; the user only needs to provide an example texture rather than writing and tuning a procedural shader.

It is desirable to have an algorithm that combines the advantages of both implicit and explicit methods. We present a new algorithm, termed *order-independent* texture synthesis, that achieves this combination. The algorithm is as general as existing explicit methods in that it can synthesize new textures simply from given examples. It is as flexible as implicit methods in that it allows textures to be evaluated on demand in any traversal order. Specifically, the algorithm allows *view-dependent* evaluation where only needed texels are synthesized on the fly, while at the same time remains *order-independent*, where the same texel always receives the same synthesized values regardless of which texels are computed and the relative traversal orders. This hybrid algorithm has numerous potential applications. For example, it can be invoked like a procedural texture shader in a ray tracing software without writing different shaders for different textures; it can be used for interactive image texture editing; it can also be implemented in graphics hardware and function like a texture mipmap for polygonal rasterization without storing the entire texture.

## 2 Previous Work

**Explicit Synthesis.** Many explicit synthesis algorithms model textures as a set of statistical features, and generate new textures by matching the statistics between the new texture and a given input sample [Heeger and Bergen 1995; Bonet 1997; Efros and Leung 1999; Wei and Levoy 2000]. These algorithms are general and can generate many different textures based on input samples. However, because these methods impose statistical dependencies between texture samples, it is impossible to evaluate only a subset of the samples while guaranteeing that these evaluated samples remain identical with respect to different synthesis orders. For example, histograms [Heeger and Bergen 1995] need to be computed from all texture samples, and pixel-neighborhood-searching [Efros and Leung 1999; Wei and Levoy 2000; Ashikhmin 2001; Tong et al. 2002] will produce different results if adjacent samples are synthesized in different orders.

Another methodology is to generate textures patch by patch [Efros and Freeman 2001; Soler et al. 2002] rather than pixel by pixel. These algorithms preserve large scale structures better than pixel-neighborhood-searching techniques. However, they usually require pre-computation of the entire patch locations or texture coordinates, regardless of the number of actual texels requested.

**Implicit Synthesis.** Existing implicit synthesis algorithms simulate the texture formation process using specialized procedures [Ebert et al. 1998]. These techniques can be highly efficient and they allow texels to be evaluated independently from each other. However, since different textures may require different procedures, these algorithms are less general than explicit methods. In addition it can be difficult to tune the parameters for those procedures to achieve the desired visual appearance of the result texture.

## 3 Algorithm

Our algorithm extends previous multi-resolution neighborhood-search texture synthesis algorithms [Wei and Levoy 2000; Tong et al. 2002]. In this section, we first describe the basic idea of our algorithm: why previous work is not order-independent, and how our approach addresses this (Section 3.1). We then describe implementation details on how to make our algorithm both fast and memory efficient (Section 3.3 and Section 3.2).

### 3.1 Basic Idea

We can summarize previous neighborhood-search algorithms [Efros and Leung 1999; Wei and Levoy 2000; Ashikhmin 2001; Tong et al. 2002] as follows.

Goal: given an input texture, generate an output texture that is similar to the input.

1. Build pyramids for both the input and output, and initialize the output.

2. Synthesize the output texels one by one from lower to higher resolutions in a certain order, such as scanline [Wei and Levoy 2000] or spiral [Efros and Leung 1999].

3. For each output texel, build a neighborhood around it. Use this neighborhood to find the best-matched neighborhood from the input, and assign the corresponding input texel to the output. Different search algorithms can be used, such as TSVQ [Wei and Levoy 2000] or coherence search [Ashikhmin 2001].

4. Repeat the above step for each output texel.

An algorithm is order-independent if we can generate texels in arbitrary order (Step 2) while guaranteeing the resulting texture to be identical, given that we start with the same input and initial conditions (Step 1). However, existing methods [Efros and Leung 1999; Wei and Levoy 2000; Ashikhmin 2001; Tong et al. 2002] are not order-independent; since their neighborhoods always include the most recently synthesized results, different traversal orders will cause different combinations of old and new values at each neighborhood, altering the result of neighborhood search. For example, consider two nearby texels, $A$ and $B$, where $A$ is within $B$'s neighborhood. If $A$ is synthesized before $B$, then the neighborhood of $B$ will include the new value of $A$; on the other hand, if $A$ is synthesized after $B$, then the neighborhood of $B$ will include the old value of $A$. As a result, the synthesized value $B$ will depend on whether or not it is synthesized before $A$.

We address this limitation by a simple idea: instead of overwriting the old values with new results, we keep the old and new values *in separate output pyramids* and use only the old values in the neighborhood search process. This idea is inspired by image convolution. When convolving image $X$ by a filter kernel to produce image $Y$, each pixel in $Y$ is computed by convolving the kernel with the "old" pixels in $X$ rather than the new pixels in $Y$. As a result, there are no dependencies among pixels in $Y$, and we can compute their values in any order.

To apply this convolution idea to texture synthesis, we keep multiple output pyramids, each storing a specific *generation* of the output. For example, we use generation 0 to store the oldest values, generation 1 to store values computed from generation 0, and so on. During synthesis, we traverse the output pyramids from lower to higher levels as usual (Step 2), but within each level, we compute the texels from lower to higher generations. To avoid dependencies between texels at the same level and generation, we modify the definition of neighborhoods (Step 3) so that, for a given output texel at a certain level and generation, its neighborhood can contain only "old" texels that are already computed, where the old texels are located at lower levels, or at the same level but lower generations. To be more precise, a texel located at ⟨level $L_i$, generation $G_i$⟩ can belong to the neighborhood of a texel located at ⟨level $L_j$, generation $G_j$⟩ only if ⟨$L_i, G_i$⟩ ≺ ⟨$L_j, G_j$⟩, where

$$\langle L_i, G_i \rangle \prec \langle L_j, G_j \rangle \text{ iff } [L_i < L_j] \text{ or } [L_i = L_j \text{ and } G_i < G_j]$$
(1)

This lexically smaller operator ≺ defines an acyclic ordering for all texels located at different levels or generations. Therefore it removes dependency between texels at the same ⟨level , generation⟩ and allows order-independent synthesis.

**Example.** Figure 1 shows an example of walking through our algorithm. The output pyramid contains 4 levels and 3 generations, and we use a 2-level neighborhood template with size 5×5. Given a user request pattern, our goal is to compute as few texels as possible to satisfy this request. From the initial pattern at ⟨$L_3, G_2$⟩ and the neighborhood templates, we can figure out the set of necessary texels at every ⟨level, generation⟩ recursively, from higher to lower levels, and higher to lower generations. For example, the footprint at ⟨$L_3, G_1$⟩ is computed by taking the union of all 5×5 neighborhoods of each texel within the "S" pattern in ⟨$L_3, G_2$⟩, and the footprint at ⟨$L_3, G_0$⟩ is determined from ⟨$L_3, G_1$⟩ in a similar fashion. After we figure out this minimum set of texels at each ⟨level , generation⟩, we can then generate them from lower to higher levels, and from lower to higher generations. That is, we start at ⟨$L_0, G_0$⟩, and then go to ⟨$L_1, G_0$⟩, ⟨$L_1, G_1$⟩, ⟨$L_1, G_2$⟩, ⟨$L_2, G_0$⟩, etc, until we end at ⟨$L_3, G_2$⟩.

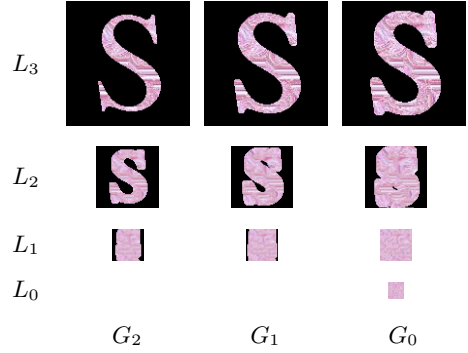

Figure 1: *Basic idea of our algorithm. In this example, we use an output pyramid with 4 levels and 3 generations, and a neighborhood template of size 5×5. ⟨$L_3, G_2$⟩ contains an user-requested pattern "S". ⟨$L_0, G_0$⟩ is initialized by randomly copying from the input. In the squares representing the remaining levels and generations, only a portion of the texels (denoted in pink above) must be computed.*

According to Equation 1, texels at ⟨$L_0, G_0$⟩ cannot use any other texels in the neighborhood; therefore they are initialized by randomly copying from the input pyramid. Similarly, texels at $G_0$ (the right-most column of images in Figure 1) can only depend on texels at lower levels since there is no texel located at even lower generations. We can consider these texels as "extrapolation" or "super-resolution" from lower pyramid levels. For the rest of ⟨level, generation⟩, the texels can depend on lower levels as well as the same level with lower generations. However, for lower level neighborhood texels, we usually use only those located at the highest generation since these are the most up-to-date values at the specific level. Note that for a fixed initial condition at ⟨$L_0, G_0$⟩, our algorithm will always generate identical results regardless of the traversal order within each ⟨level, generation⟩. This is similar to Perlin noise where the computed texel values remain invariant given a fixed random permutation table.

### 3.2 Making It Memory Efficient: Texture Cache

Our basic algorithm presented above is not memory efficient since it keeps multiple generations of the output. Fortunately, for many applications where only a subset of the texture is accessed (Figure 1), we only need to store this subset. In addition, since most practical rendering algorithms exhibit good texture coherence [Hakura and Gupta 1997], we can further reduce the storage requirement by caching only the recently computed/accessed values.

Our revised algorithm works as follows. Instead of storing the entire output, we use a small texture cache to store already-generated texels. This is different from prior methods [Hakura and Gupta 1997] where the cache is only used to reduce texture memory reads. In the beginning of the algorithm, the cache is empty and every requested texel needs to be computed. However, as the cache gradually fills up, previously computed texels may be requested again, and they can be found in the cache without any computation. If a cached texel is kicked out of the cache due to capacity limit, we may need to re-compute it if it is requested again. The actual extent of re-computation depends on the rendering algorithm that drives these requests. Fortunately, since most applications exhibit good coherence, the ratio of re-computation is usually low, as we shall see in the result section (Section 4).

### 3.3 Making It Fast: K-Coherence Search

Our algorithm is orthogonal with respect to the specific neighborhood-search algorithm used in Step 3. However, since the search is in the inner loop, the choice of the search procedure will determine the overall speed of our algorithm. In addition, the search procedure will determine the quality of texture synthesis results.

We have experimented with several different approaches and found that the K-coherence algorithm [Tong et al. 2002] provides the best quality/speed trade off. The algorithm has constant time complexity per output texel and provides state-of-art synthesis quality. The algorithm is divided into two phases: analysis and synthesis. During analysis, the algorithm builds a similarity-set for each input texel, where the similarity-set contains a list of other texels with similar neighborhoods to the specific input texel. During synthesis, the algorithm builds a candidate-set by taking the union of all similarity-sets of the neighborhood texels for each output texel, and then searches through this candidate-set to find out the best match. The size of the similarity-set, $K$, is a user-controllable parameter that determines the overall speed/quality. The authors in [Tong et al. 2002] have found that a small $K$ (in the range of 1 to 11) works well for a variety of textures.

### 3.4 Summary

We can summarize our algorithm as follows. Given an input texture, an output texture cache size, a neighborhood template, and a texture access pattern:

1. Build a Gaussian pyramid for the input. Pre-process the input for the specific search algorithm, if necessary. The output pyramid is implemented as a cache, capable of holding texels at multiple generations. The lowest ⟨level, generation⟩ is initialized by randomly copying from the input.

2. For each requsted texel, determine the minimum set of neighborhood texels located at lower levels and generations. This can be achieved by recursively expanding the request patterns from higher to lower levels, and from higher to lower generations, as shown in Figure 1.

3. For texels in this minimum set that are not in the cache (including the requested texel), synthesize them one by one from lower to higher levels, and from lower to higher generations. The synthesis can be done via any neighborhood-search algorithms; we use K-coherence search in this paper.

4. Repeat the above two steps for each requsted texel that is not in the cache.

## 4 Results

In this section, we demonstrate several aspects of our algorithm's performance. We begin by demonstrating the effects of various synthesis parameters on image quality. We then characterize the cache performance under different access patterns.

**Synthesis Quality**

We have applied our approach with many different textures, and we found that it works as well as the original K-coherence search algorithm [Tong et al. 2002]. The only additional parameter we need to set is the number of generations. We have found that 2 or 3

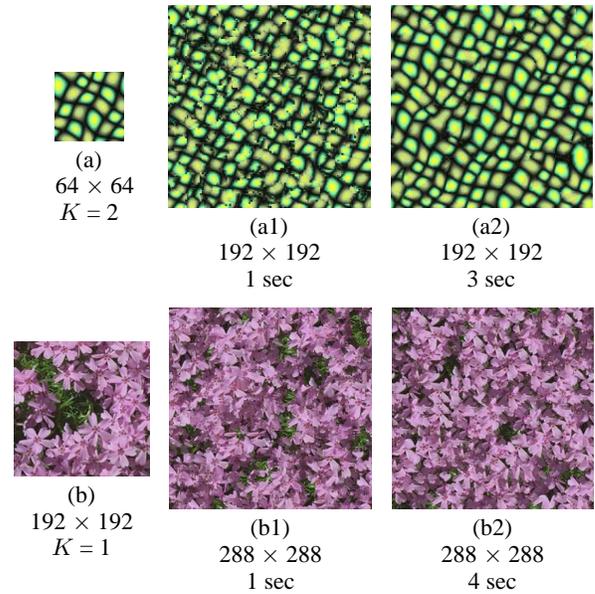

(a)
64 × 64
K = 2

(a1)
192 × 192
1 sec

(a2)
192 × 192
3 sec

(b)
192 × 192
K = 1

(b1)
288 × 288
1 sec

(b2)
288 × 288
4 sec

Figure 2: *Synthesis Quality. For each group of images, the input sample is on the left, the result with 1 generation is shown in the middle, and the result with 3 generations is shown on the right. Below each figure, we label the image size and computation time in seconds, measured on a 1.1GHz AMD Athlon PC. These synthesis results compare favorably to the best existing algorithms.*

generations work well for all textures we tried. If only 1 generation is used, the algorithm essentially uses only lower-resolution information for synthesis. This is very similar to [Bonet 1997] and, as shown in Figure 2, the result textures have similar artifacts.

**Cache Performance**

As in most cache-based algorithms, the size of the cache plays a critical role in its performance. When the cache in our algorithm is large, it can hold all the computed texels. However when the cache is small, it cannot hold all texels, and some texels may be computed multiple times. As a result, the performance of small caches is affected by the coherence of the request patterns.

We analyze the performance of our algorithm under different cache sizes using ray tracing and polygonal rasterizer scenes. We collect texel requests from the renderer of these scenes, and feed them into our algorithm for simulation. In our experiment, we use 3 benchmarks with different texture characteristics and triangle sizes, as shown in Figure 3:

- **Teapot.** This is a ray-casting scene containing a large textured teapot occluded by 3 non-textured teapots. The texture has size 256×256; due to ray-casting, the occluded portion of the texture is not computed and has the shape of the projected occluders. The scene is rendered by tracing rays in 32-pixel-wide vertical swathes, from top to bottom, left to right.

- **Single Polygon.** This scene contains a large textured polygon with size 512×512, viewed in perspective. The polygon is tessellated into 64×64 tiles and covered by a large 512×512 texture pyramid with 4 levels. Since the polygon is clipped by the viewing frustum, only 19 percent of the texels are requested by the rasterizer, and this portion is wedge-shaped due to the perspective view. Our algorithm synthesizes 23 percent of all cache pixels. Although this is slightly larger than 19 percent, it is still 4 times faster than synthesizing the whole texture using K-coherence search.

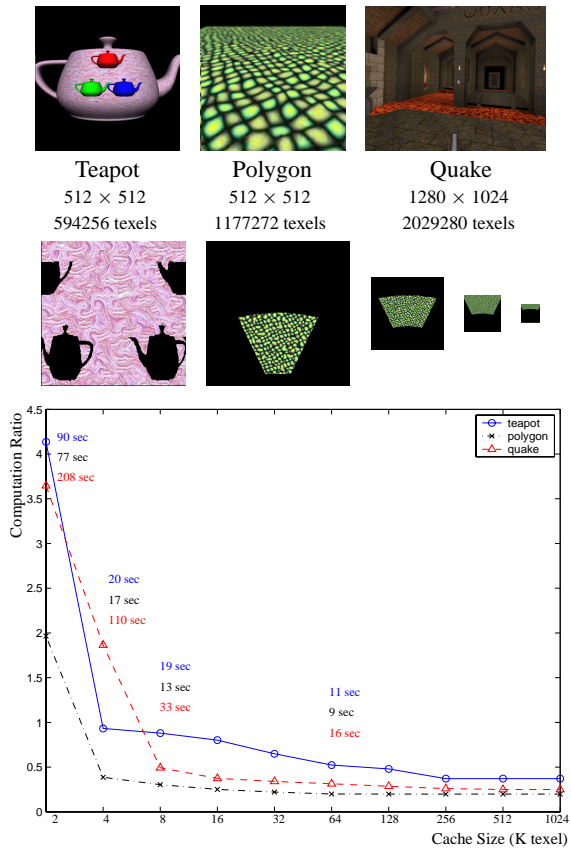

**Teapot**
512 × 512
594256 texels

**Polygon**
512 × 512
1177272 texels

**Quake**
1280 × 1024
2029280 texels

Figure 3: *Texture cache performance for several scenes: teapot, polygon, and quake. (Top) The screen shots of the 3 scenes, along with the frame size and total number of requested texels. (Middle) The requested texels for the teapot (only the highest resolution) and polygon (all 4 resolutions) scenes. The black regions indicate texels not requested (and therefore not synthesized). (Bottom) The cache size v.s. performance. We measure the performance in terms of computation ratio, which is the average number of computed texels per requested texel. The cache replacement policy is LRU. We also mark wall-clock simulation time (synthesis + cache emulation) at several key cache sizes.*

- **Quake.** This scene shows a frame from the OpenGL port of the video game Quake, containing mainly architectural walk-throughs with large polygons. We choose this benchmark to see how rasterizing large polygons may degrade cache coherence. The original Quake game uses many small textures to reduce texture load; although this works well for regular patterns such as brick walls, small textures may introduce unnatural visual repetitions for stochastic textures such as the lava. Our algorithm can address this problem by synthesizing large textures on the fly from a small input. We demonstrate this by synthesizing the lava texture with size 512×512 from a small 64×64 crop, and eliminate the majority of the unnatural repetitions in the rendering.

We measure the performance of our algorithm by *computation ratio*, which we define to be the average number of synthesized texels per requested texel. The performance is better when the computation ratio is lower since less computation is involved. In Figure 3, we plot the computation ratio versus different cache sizes for all three benchmarks. The algorithm performs reasonably well at small cache sizes (between 2K to 8K), and the computation ratio drops as the cache size increases. The performance remains roughly constant after the cache size reaches 8K, indicating that a small cache size is sufficient to hold the working set for these scenes.

**Computation Time**

As shown in Figure 2, our current implementation takes a few seconds to generate typical textures on a commodity PC without any hardware acceleration. This is sufficiently fast for interactive applications with small windows (such as the ray tracer shown in the video). However, for full-screen applications such as the Quake game in Figure 3, our current implementation is not fast enough. We plan to address this issue by implementing our algorithm on a programmable graphics hardware. This is likely to provide at least 2 orders of magnitude speed-up over our software-only implementation.

# 5 Conclusions and Future Work

In this paper, we have presented and analyzed a new algorithm for order-independent texture synthesis. The algorithm allows texture samples to be generated in arbitrary order on demand, yet the resulting texture always looks the same. The algorithm has comparable image quality with the state of art algorithms. It is computationally efficient with K-coherence search. It is storage efficient due to its use of a texture cache. We demonstrate that small caches are sufficient by analyzing our algorithm through different texture mapping scenes.

There are several possible directions for future work. Although our algorithm is fast enough for software applications such as ray tracing and image editing, it needs furthur speed improvements to be practical for hardware polygonal rasterizers. This can be achieved by finding a search algorithm more efficient than K-coherence search. Another possibility is to implement our algorithm as a fragment program in new generation GPUs. The fragment program would synthesize the texture in multiple passes, where each pass renders a specific generation/resolution of the texture. Our algorithm can also be used as a texture decompresser for a software viewer such as VRML or QuickTime VR. This would substantially reduce the storage space and transmission time for viewing scenes containing large textured regions. Finally, our algorithm can be implemented in a shading language and integrated with a ray tracing package. This could make statistical texture synthesis more useful to animators, who are more accustomed to procedural shaders.